# A note on the constant characteristic time of failure incubation processes under various high-rate loads

*Ivan Smirnov*

*Saint Petersburg State University, Universitetskaya nab. 7/9, St. Petersburg, 199034, Russia*

Corresponding author: ivansmirnov.sci@gmail.com

**Abstract.** The research reveals the existence of a constant characteristic time of preparatory micro-structural processes before the onset of macro-failure at various high loading rates of brittle and quasi-brittle materials. The presence of this characteristic is analysed based on available data in the literature from dynamic tests for uniaxial compression and splitting. It is shown that the characteristic time can be determined experimentally and used to calculate the strain rate dependencies of either critical failure stresses or time to failure, at least in the case of linearly growing loads. In addition, it is discussed that the presence of this constant parameter opens up a prospective opportunity for research and development of new methods for assessing the structural-temporal and scale characteristics of the strength and failure of materials under dynamic loads.

**Keywords:** quasi-brittle material, high strain rate, dynamic strength, time to fracture, characteristic time, failure criteria.

## 1. Introduction

The problems of deformation and failure of materials at high-rate intensive loads have been relevant for a significant period of time. Today, these studies continue to both certify the behaviour of materials at a given range of strain rates [1–3], and solve fundamental issues in order to develop approaches for qualitative forecasting the behaviour of materials, regardless of loading history [1,4,5]. However, there are still no universally accepted test methods or parameters that could allow us to evaluate and model the dynamic strength of materials based on simple engineering principles.

At sufficiently slow loads, the tensile strength and yield strength can, in fact, be assumed to be constant and considered as characteristic strength parameters of a material. This forms the basis of systems for state and industry standards to determine the strength characteristics of brittle and plastic materials. When the strain or loading rates exceed the standardized quasi-static range, these strength characteristics can become unstable and often strongly depend on the load history [5,6]. Therefore, under dynamic loads, they can no longer be considered as material parameters. To date, there is no agreement on which characteristics can serve as the primary indicators of the properties of materials under high-rate and shock loads. There is an urgent need to identify the characteristics, which do not depend on the history and method (set-up) of load application.

This short communication presents an important, previously unknown effect of the connection between failure incubation processes in the material structure and a constant time characteristic, which are independent of strain/loading rate. It is shown that this characteristic time can be determined experimentally and applied to criteria for evaluating the strength of a material within a wide range of loading rates.

**2. The characteristic time of failure incubation processes**

Let us consider the case of uniaxial loading for which failure begins at the stage of load growth without the pronounced and irreversible deformation of a specimen. This situation is typical when testing brittle and quasi-brittle materials, using the scheme of uniaxial compression or splitting tensile tests, for example. Let the beginning of a decrease in stresses in the stress-time or stress-strain diagram be the factor indicating the beginning of failure. The strength of a material then corresponds to the maximum value of stress. This is a common procedure for determining the strength of such materials with quasi-static tests. However, with an increase in the load or strain rate above the quasi-static range, the maximum failure stresses increase. Further in this paper, the maximum failure stresses are assumed as critical stresses.

Fig. 1 demonstrates the case described above. Since the loading area up until the moment of failure can be considered linear, the next expression is valid for the critical stress $\sigma_{cr}$:

$$\sigma_{cr} = E\dot{\varepsilon}t_{cr}, \quad t_{cr} > 0, \tag{1}$$

where $E$ is the modulus of elasticity, $\dot{\varepsilon}$ is the strain rate and $t_{cr}$ is the time before the start of failure. When the critical stress $\sigma_{cr}$ is greater than the quasi-static strength $\sigma_{st}$, the time to failure and expression (1) can be written as the sum of two terms:

$$t_{cr} = t_{st} + t_d = \frac{\sigma_{st}}{E\dot{\varepsilon}} + t_d, \tag{2}$$

$$\sigma_{cr} = \sigma_{st} + E\dot{\varepsilon}t_d, \quad t_{st} + t_d > 0, \tag{3}$$

where $t_d$ is the time from time $t_{st}$ when the quasi-static strength of a material is reached to the time of the actual critical stress $t_{cr}$. As a rule, experimental equipment allows for recording time diagrams of load and strain. The strain or loading rate is specified by the input test conditions. For the case under consideration, the stress rate is $\dot{\sigma} = E\dot{\varepsilon}$. Thus, all components of expression (3) can be determined from the experiment.

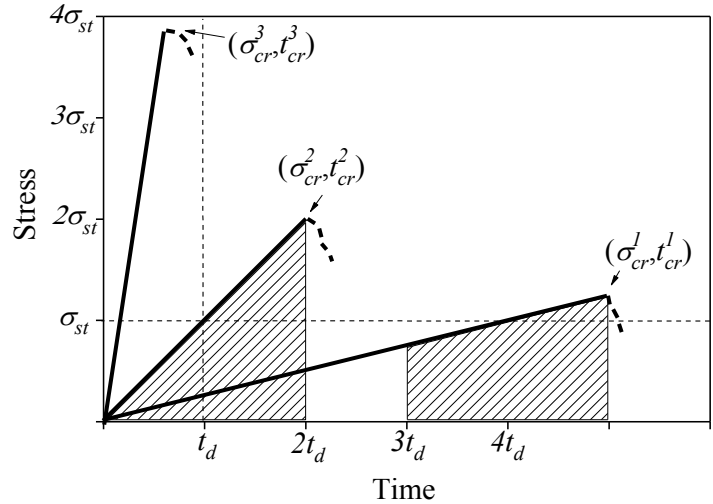

**Fig. 1.** Schematic representation of loading at different rates. Designations are disclosed in the text.

Then we can estimate the time $t_d$ that characterises the dynamics of micro-processes in the structure of the material after reaching the quasi-static strength of the material $\sigma_{st}$ until the onset of 'macro' failure $t_{cr}$:

$$t_d = \frac{\sigma_{cr} - \sigma_{st}}{E\dot{\varepsilon}}. \tag{4}$$

Let us consider such an estimate regarding various experimental data. Dynamic strength is typically evaluated by the dependence of critical stress on the strain or stress rate. An example of such dependence for the case of dynamic uniaxial compression of concrete is presented in Fig. 2. Substituting the experimental points in expression (4), we derive the dependence of time $t_d$ on the strain rate. Fig. 3 presents such calculations for the uniaxial compression and splitting test of various materials. Usually experimental points have a spread, so it is convenient to use the equation of the slope of a regression line to determine the time $t_d$. The results show that, regardless of the material and loading scheme, the time $t_d$ for each experiment can be considered constant.

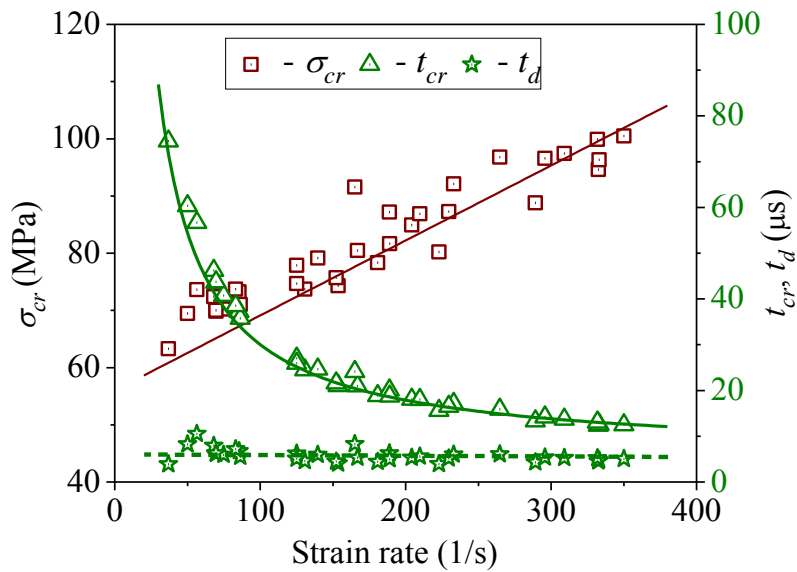

**Fig. 2.** Strain rate dependencies of critical stresses and time to failure for dynamic uniaxial compression testing of concrete. Experimental data were obtained by [7]. The solid curves ware calculated using expressions (2) and (3). The dashed line is a linear approximation.

Note that we use the same notation for the case of compression and splitting. This is not only to avoid unnecessary notation, but also to show the commonality of the effect for various tests. A caveat, however, is that when considering a specific test method, these load and material parameters only apply to this method of testing.

Up to this point, it has been assumed that the time to failure $t_{cr}$ is greater than the time $t_d$. However, the question arises: up to what value does the time to failure decrease, and would it be less than time $t_d$? In order to answer this question, it is necessary to consider the stress impulse before the start of failure.

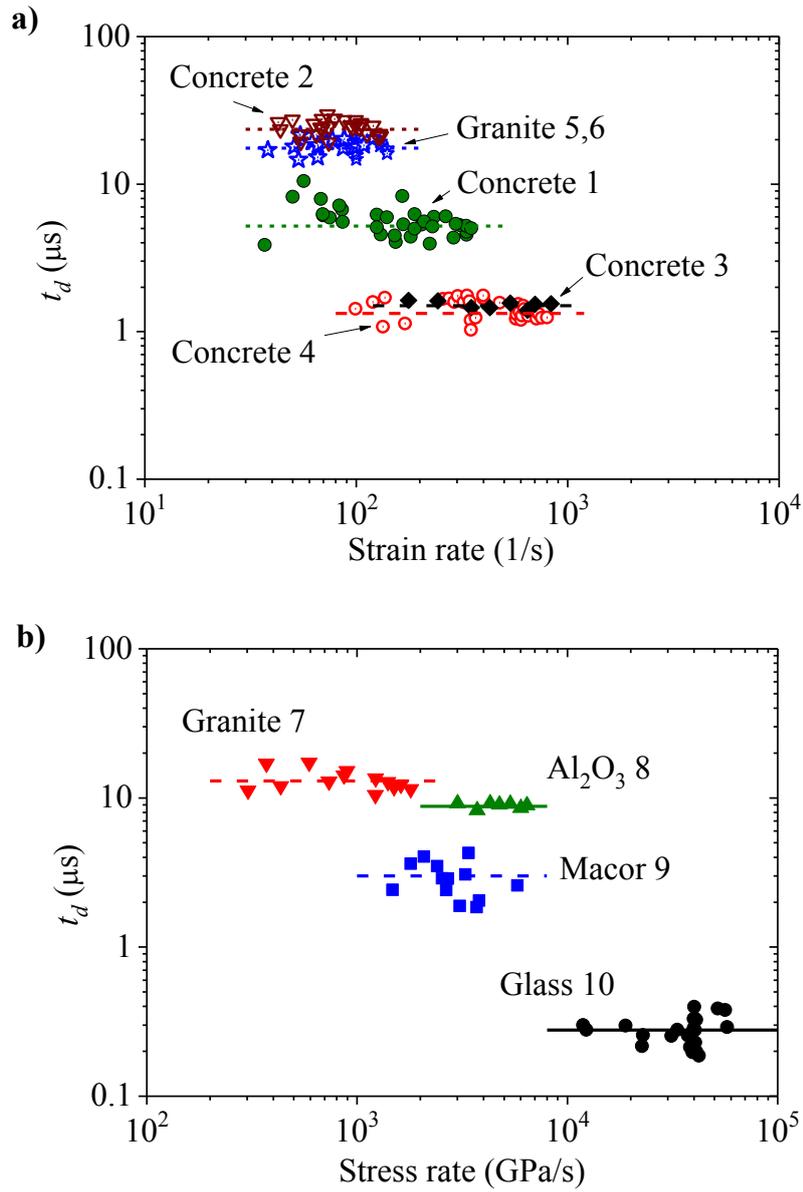

**Fig. 3.** The characteristic times of incubation micro-processes from the moment of reaching quasi-static strength to the moment of the onset of macro-failure of various materials under high-rate loading. The calculations were carried out by Eq. (4) for the experimental data: a) compression tests of concrete 1 [7], 2 [8], 3 [9], 4 [10], and granite 5 [11] and 6 [12]; b) splitting test of granite 7 [13], ceramics 8 [14] and 9 [15] and glass 10 [16].

A stress analysis shows that for critical stress $\sigma_{st} < \sigma_{cr} < 2\sigma_{st}$, the stress impulse applied to a specimen is constant over a time from $t_{cr} - 2t_d$ to $t_{cr}$ (at the condition that $t_d$ is constant). These segments of the stress pulses $J_{cr}$ are indicated by the shaded areas in Fig. 1. It is evident that the stress impulse $J_{cr}$ is equal to

$$J_{cr} = 2\sigma_{st} t_d. \tag{5}$$

Suppose this impulse is the minimum value of the stress impulse that must be applied to the specimen in order to initiate and prepare its macro failure. Then, if the time to failure is $t_{cr} \leq 2t_d$, the failure stress impulse must also be at least $2\sigma_{st}t_d$. Since for $t_{cr} \leq 2t_d$ $2\sigma_{st}t_d = 0.5\sigma_{cr}t_{cr}$, then the expressions for the strain rate dependence of the critical stress and time to failure, taking (1) into account, can be written in the following form:

$$\sigma_{cr} = \sqrt{4E\sigma_{st}t_d\dot{\varepsilon}}, \quad t_{cr} \leq 2t_d, \qquad (6)$$

$$t_{cr} = \sqrt{\frac{4\sigma_{st}t_d}{E\dot{\varepsilon}}}, \quad t_{cr} \leq 2t_d. \qquad (7)$$

In addition, the condition for time to failure in expression (3) should now be specified as $t_{cr} \geq 2t_d$.

The characteristic time $t_d$ can be easily obtained from expression (6). Fig. 4 and 5 present these calculations for the uniaxial compression and splitting test of various materials. Thereby, the time $t_d$ was determined by formula (4) for the case $\sigma_{cr} < 2\sigma_{st}$ and from formula (6) for $\sigma_{cr} \geq 2\sigma_{st}$. This example also demonstrates that the time $t_d$ can be considered constant for various strain and loading rates.

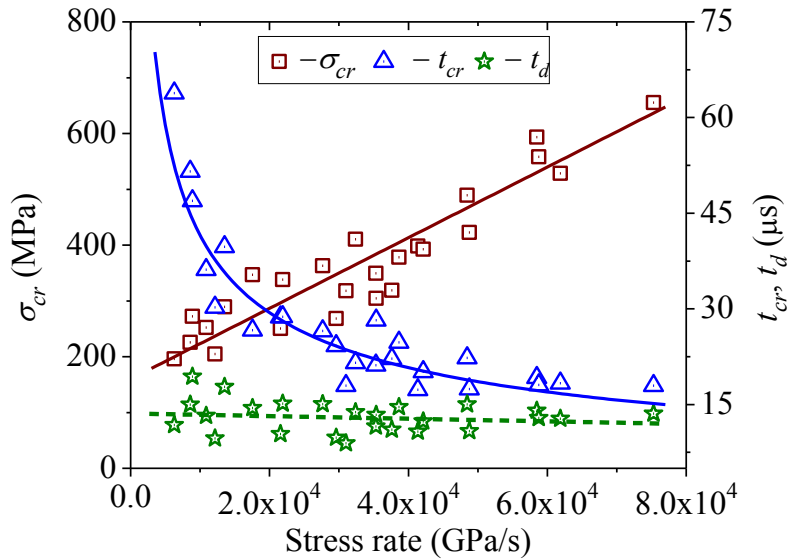

**Fig. 4.** Strain rate dependencies of critical stresses and time to failure for dynamic uniaxial compression tests of concrete in the cases of $t_{cr} \geq 2t_d$ and $t_{cr} \leq 2t_d$. Experimental data was obtained by [17]. The solid curves were calculated using conditions (2), (3), (6) and (7). The dashed line is a linear approximation.

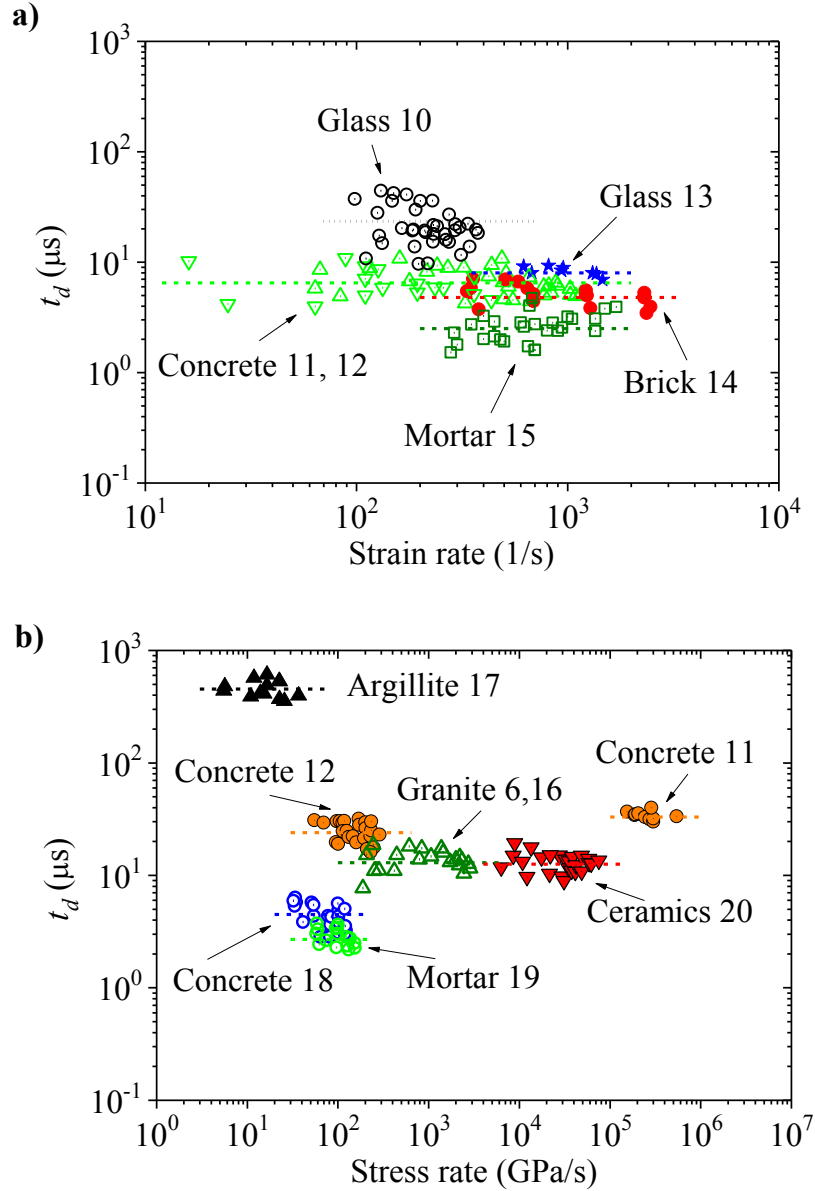

**Fig. 5.** The characteristic times calculated for different conditions of $\sigma_{cr} < 2\sigma_{st}$ and $\sigma_{cr} \geq 2\sigma_{st}$. The calculations were carried out according to Eq. (4) and (6) for the experimental data: a) compression tests of concrete 11 [18] and 12 [19], glass 10 [16] and 13 [20], brick 14 [21], and mortar 15 [22]; b) splitting test of rocks 6 [12], 16 [23], and 17 [24], concrete 11 [18] and 12 [19], and 18 [25], mortar 19 [25], and ceramics 20 [26].

## 3. Discussion and future work

Thus, the experimental data provides reason to introduce the characteristic time of incubation processes of failure $t_d$, which can be considered a constant value independent of a strain rate above the quasi-static range, at least for cases of continuous linear increase in load during the uniaxial compression or splitting tests. This time parameter characterises the individual response of a material to dynamic loads. Therefore, this characteristic time can be used in criteria conditions of structural-temporal approaches to failure analysis and strength prediction. For example, similar expressions for (3) and (6) were obtained theoretically using the incubation time criterion [5,27,28]. The criterion assumes that the following condition must be implemented for failure to occur:

$$\int_{t-\tau}^{t} \sigma(t)dt \geq \sigma_{st}\tau, \qquad (8)$$

where $\sigma(t)$ is the stress profile at the failure place, $\sigma_{st}$ is the quasi-static strength, and $\tau$ is the incubation time of failure. The parameters $\sigma_{st}$ and $\tau$ are strength parameters of a material for a particular test scheme, for example, compression or tension. The incubation time was introduced as a hypothetical characteristic period of time responsible for the incubation period of macro-failure. This was necessary to ensure the possibility of a smooth transition from pulsed loads to quasi-static loads using the Nikiforovsky-Shemyakin integral criterion for spall fracture (full integral of the stress over time at the spall section should reach a critical value) [5,29]. Therefore, according to (8), failure will occur in the case that the stress impulse in the region of failure is not less than $\sigma_{st}\tau$ for the time $t \leq \tau$ (for $t < 0$, $\sigma(t) = 0$). A similar assumption is made in the discussion of (5).

Substituting (1) for $\sigma(t)$ in (8), we can obtain simple relationships for calculating the strain rate dependencies of critical stresses $\sigma_{cr}$ and the time to failure $t_{cr}$:

$$\begin{cases} \sigma_{cr} = \sigma_{st} + 0.5E\dot{\varepsilon}\tau, & t_{cr} = \dfrac{\tau}{2} + \dfrac{\sigma_{st}}{E\dot{\varepsilon}}, & t_{ct} > \tau, \\ \sigma_{cr} = \sqrt{2E\sigma_{st}\tau\dot{\varepsilon}}, & t_{cr} = \sqrt{\dfrac{4\sigma_{st}t_d}{E\dot{\varepsilon}}}, & t_{cr} \leq \tau. \end{cases} \qquad (9)$$

Substituting $\tau = 2t_d$, we derive the expressions in (9), which correspond exactly to expressions (2), (3), (6) and (7).

The incubation time approach allows us to successfully solve a number of dynamic problems related to fracture mechanics [5,27–30]. However, the question of experimental determination of the incubation time of failure still remains open-ended. The obtained result shows that the incubation time of failure, at least for the case of a controlled linear increase in fast loading, can be determined experimentally by estimating $t_d$.

The demonstrated effect opens up new possibilities for solving important problems related to the mechanics of dynamic deformation and fracture of materials. One of these tasks relates to the possibility of determining the parameters of the dynamic strength of materials using basic tests that are both publicly available and generally accepted. Furthermore, it is necessary to study appropriate methods and basic rules to determine the characteristic time $t_d$. Experimental conditions affecting time $t_d$ should also be established.

Another problem relates to the determination of the strain rate dependence of the yield strength. The presented reasoning can be applied to consider similar characteristic time of incubation processes involved in the plastic deformation [30].

The found characteristic time of failure incubation processes can provide a new approach to the determination of scale levels of failure. Since we have a constant time characteristic for the implementation of preparatory failure processes, we can assume that there is also a constant characteristic structural scale at which these processes are realised. For example, according to the structural-temporal approach [29], this characteristic scale can relate to the propagation distance of the elastic wave during time $\tau$ ($d = \tau C$, in which $C$ is the speed of sound); as well, the scale can be introduced as $d = (2KI_C^2)/(\pi\sigma_{st}^2)$ ($KI_C$ as the stress intensity factor). However, the accuracy of these expressions for the elementary scale of failure is still being studied. Thus, the relationship of time $t_d$, scale factors and the dynamics of the preparatory microstructural processes of macro-fracture should be studied.

## 4. Conclusions

The presented study shows the existence of a characteristic time for preparatory micro-processes of macro-failure of brittle and quasi-brittle materials. This time does not depend on the strain or loading rate, at least in terms of the compression and splitting test methods under consideration. The value of the characteristic time can be determined directly from the experiment.

The presented results show that this integral time characteristic of the dynamic failure process can be used for the development of experimental and theoretical foundations to determine and predict the strength characteristics of constructional materials across a wide range of loading rates.

**Acknowledgements**

This work was supported by the Russian Science Foundation, grant № 18-79-00193.